\documentclass[sigconf,screen]{acmart}

\usepackage{hyperref}
\usepackage{hyperxmp}
\usepackage{amsmath}
\usepackage{graphicx}
\usepackage{multirow}
\usepackage{pifont}
\usepackage{tabularx}
\usepackage{booktabs}
\usepackage{tablefootnote}
\usepackage{booktabs}
\usepackage{array}
\usepackage{balance}
\usepackage{multirow}
\usepackage{color}
\definecolor{lightgray}{RGB}{215,215,215}
\usepackage{colortbl}  
\usepackage{xcolor}
\usepackage{subfigure}
\usepackage{algorithm}  
\usepackage{algorithmicx}
\usepackage[noend]{algpseudocode}
\usepackage{amsmath}  
\usepackage{tabularx}
\usepackage[utf8]{inputenc}
\usepackage[english]{babel}
\usepackage{amsthm}
\usepackage{bm}
\usepackage{acronym}
\usepackage{lipsum}   
\usepackage{xcolor}
\usepackage{booktabs}
\usepackage[inline]{enumitem}
\usepackage{makecell}
\usepackage[skip=2pt]{caption}
\usepackage{adjustbox}

\author{Zhen Zhang}
\authornote{These authors contributed equally to this work.}
\authornote{Zhen Zhang and Xin Xin are affiliated with Shandong University; the Shandong Provincial Key Laboratory of Computing-Network Integration, Shandong University; and the Qingdao Key Laboratory of Trustworthy Artificial Intelligence, Shandong University.}
\orcid{0009-0001-0290-5386}
\affiliation{%
  \institution{Shandong University}
  \city{Qingdao}
  \country{China}
}
\email{zhen.zhang.sdu@gmail.com}

\author{Jujia Zhao}
\authornotemark[1]
\orcid{0009-0003-6951-7593}
\affiliation{%
  \institution{Leiden University}
  \city{Leiden}
  \country{The Netherlands}
}
\email{zhao.jujia.0913@gmail.com}

\author{Xinyu Ma}
\orcid{0000-0002-5511-9370}
\affiliation{%
  \institution{Baidu Inc.}
  \city{Beijing}
  \country{China}
}
\email{xinyuma2016@gmail.com}

\author{Xin Xin}
\orcid{0000-0003-4703-7356}
\authornotemark[2]
\affiliation{%
  \institution{Shandong University}
  \city{Qingdao}
  \country{China}
}
\email{xinxin@sdu.edu.cn}

\author{Maarten de Rijke}
\orcid{0000-0002-1086-0202}
\affiliation{%
  \institution{University of Amsterdam}
  \city{Amsterdam}
  \country{The Netherlands}
}
\email{m.derijke@uva.nl}

\author{Zhaochun Ren}
\authornote{Corresponding author.}
\orcid{0000-0002-9076-6565}
\affiliation{%
  \institution{Leiden University}
  \city{Leiden}
  \country{The Netherlands}
}
\email{z.ren@liacs.leidenuniv.nl}

\renewcommand{\shortauthors}{Zhen Zhang et al.}

\copyrightyear{2026}
\acmYear{2026}
\setcopyright{cc}
\setcctype{by}
\acmConference[SIGIR-AP 2026]{Proceedings of the 2026 Annual International ACM SIGIR Conference on Research and Development in Information Retrieval in the Asia Pacific Region}{December 13--15, 2026}{Singapore, Singapore}
\acmBooktitle{Proceedings of the 2026 Annual International ACM SIGIR Conference on Research and Development in Information Retrieval in the Asia Pacific Region (SIGIR-AP 2026), December 13--15, 2026, Singapore, Singapore}
\acmDOI{10.1145/3848000.3850210}
\acmISBN{979-8-4007-3033-7/2026/12}


\title[A Reproducibility Study of Cold-Start Recommendation in Generative Recommenders]{A Reproducibility Study of Cold-Start Recommendation\\ in Generative Recommenders}

\begin{document}

\begin{abstract}
Recommender systems constantly face the cold-start challenge, requiring them to serve new users and newly introduced items with limited or no historical interaction signals. 
Recently, generative recommenders built on pre-trained language models have been increasingly expected to mitigate this issue by leveraging semantic knowledge and reasoning over sparse contexts. 
However, existing studies rarely evaluate cold-start as a primary setting, and their reported results are often difficult to attribute because model scale, identifier design, and training strategy are changed simultaneously.

In this work, we present a controlled reproducibility study of representative generative recommenders under unified user and item cold-start protocols across three benchmark datasets. 
By systematically isolating key design dimensions, we uncover four critical insights. 
First, the cold-start challenge exhibits strong asymmetry: while generative models maintain relatively stable performance when serving new users with sparse histories, their accuracy degrades sharply when attempting to recommend completely unseen items. 
Second, scaling up the underlying language model provides only marginal gains and cannot fundamentally close this cold-start performance gap. 
Third, ID design dictates strict performance trade-offs: textual IDs significantly improve item cold-start at the expense of warm-start accuracy, whereas semantic IDs are robust overall but fail severely on unseen items. 
Our analysis reveals that this failure stems from auto-regressive constrained decoding, which amplifies structural biases toward frequently observed branches during training. 
Finally, reinforcement learning optimized on seen data fails to consistently improve cold-start performance. 
Our study provides reproducible evidence and practical insights for enhancing the generalization capabilities of generative recommendation systems in cold-start scenarios.\footnote{Our data and code are available at \url{https://github.com/zhangzhen-research/ColdGenrec}.
\label{fn:1}}

\end{abstract}


\maketitle
\section{Introduction}
\label{sec:introduction}

Recommender systems are a core component of two-sided marketplaces, matching users with items in dynamic, open-world environments~\cite{fu2026differentiable,lin2025order}.
A fundamental challenge in these settings is cold-start on both sides: recommending relevant items to newly registered users (\emph{user cold-start}) and exposing newly introduced items to existing users (\emph{item cold-start})~\cite{wei2021contrastive,huang2025large}.  
In these data-sparse scenarios, initial recommendations dictate long-term user retention, while early exposure determines whether new items can acquire the feedback necessary for model convergence.
Delivering effective recommendations under user and item cold-start conditions remains a central challenge for building robust recommender systems.

\begin{figure}[h]  
\setlength{\abovecaptionskip}{0cm}
\setlength{\belowcaptionskip}{0cm}
    \centering    
    \includegraphics[width=\linewidth]{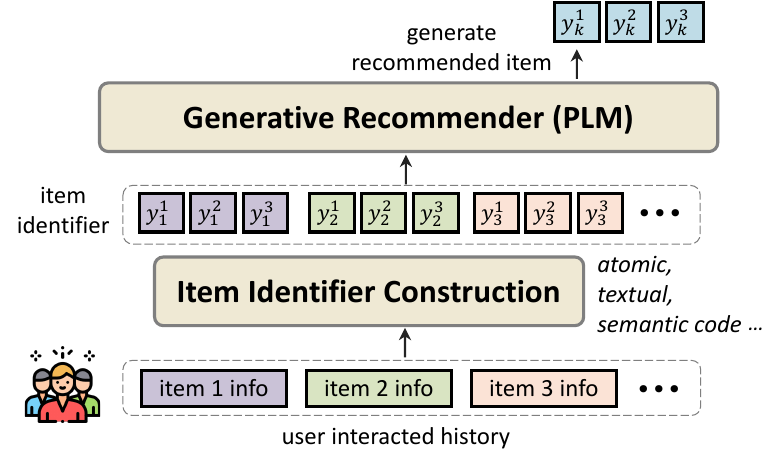}
    \caption{Generative recommendation pipeline.}
    \label{fig:intro}
\end{figure}

Most traditional recommender models rely heavily on historical interactions and therefore perform poorly when either users or items lack reliable feedback signals~\cite{ding2024inductive}.
To address cold-start, these methods typically incorporate auxiliary mechanisms, such as using side information, data augmentation, or meta-learning, to construct or adapt representations for new users and items~\cite{monteil2024marec,zhang2022diverse,vartak2017meta}.
More recently, generative recommendation built on pre-trained language models (PLMs) has emerged as an alternative paradigm and is often expected to mitigate cold-start for both the user and item sides~\cite{zhang2025cold}.
Figure~\ref{fig:intro} illustrates a general pipeline of generative recommendation.
For user cold-start, PLMs can use test-time conditioning to infer user preferences from limited interaction context, thereby enabling generalization to previously unseen or sparsely observed users.
For item cold-start, pre-trained knowledge can help interpret the semantic content of newly introduced items through their identifiers (e.g., textual descriptors or semantic codes)~\cite{tan2024idgenrec, rajput2023recommender}.

Despite these hypotheses, cold-start is seldom treated as a primary evaluation setting for generative recommendation, leaving open how these models behave when users or items have little to no interaction history. 
Moreover, existing generative recommendation methods often conflate multiple design factors, such as model scale, identifier design, and training strategy, which makes factor-wise attribution difficult. 
Even when performance differences are observed, it remains unclear which factors are responsible.

This gap motivates a systematic reproducibility study that isolates and evaluates the effects of key confounding factors under a unified set of cold-start settings. 
Our reproducibility study is guided by four research questions:
\begin{enumerate}[leftmargin=*,label=\textbf{(RQ\arabic*)}]
    \item How do representative generative recommenders perform under cold-start settings?
    \item How does model scale influence cold-start performance in generative recommendation?
    \item How does identifier design (e.g., atomic, semantic identifiers) affect the ability of generative models to generalize to unseen users and items?
    \item How do different training strategies (e.g., supervised fine-tuning, or reinforcement learning) impact cold-start behavior?
\end{enumerate}
Since identifier design emerges as the most influential factor for item cold-start performance, we further conduct a focused analysis of the decoding behavior of semantic identifiers to better understand why they fail on unseen items.

\smallskip\noindent\textbf{Findings.}
Our experiments provide the following key findings in response to the above research questions:
\begin{enumerate*}[label=(\roman*)]
\item The cold-start problem exhibits highly asymmetric behavior: while user cold-start leads to only moderate performance degradation, item cold-start remains substantially more challenging for existing generative recommendation approaches. 
\item Increasing model size yields only marginal gains under cold-start settings and fails to close the massive performance gap between warm and item cold-tests, indicating that scaling alone is insufficient.
\item Identifier design plays a critical role and reveals a clear trade-off.
Textual identifiers improve item cold-start generalization but hurt warm-start and user cold-start accuracy.
Semantic identifiers are more robust overall, but their constrained decoding space biases generation toward training-observed branches, limiting new-item generalization.
\item Reinforcement learning does not consistently improve cold-start performance and may even reduce robustness under distribution shift. 
This suggests that reward objectives optimized on seen interactions may reinforce in-distribution patterns rather than improve generalization to unseen users and items, highlighting the need for cold-start-aware reward design.

\end{enumerate*}


\smallskip\noindent\textbf{Contributions.}
Our main contributions are as follows:
\begin{enumerate*}[label=(\roman*)]
\item We conduct a comprehensive evaluation of representative generative recommenders under a unified suite of user and item cold-start settings, providing a controlled assessment of their cold-start behavior.
\item We establish a reproducible cold-start evaluation framework that isolates key design factors, including model scale, identifier design, and training strategy, enabling systematic and factor-wise analysis of generative recommenders under controlled protocols.
\item We analyze the failure mode of semantic identifiers under item cold-starts; their weakness is closely tied to position-wise constrained decoding and training-observed branch bias.
\item We offer actionable insights for enhancing the generalization capabilities of generative recommendation systems in cold-start scenarios.
\end{enumerate*}

\section{Related Work}
\label{sec:related-work}

\subsection{Generative retrieval} 
With the rapid advancement of large language models (LLMs)~\cite{vaswani2017attention, zhao2023survey}, generative retrieval has emerged as a paradigm shift in information retrieval. 
Unlike the traditional ``index-retrieve-rank'' pipeline that relies on sparse or dense vector matching (e.g., BM25, DPR)~\cite{robertson2009probabilistic, karpukhin2020dense}, generative retrieval unifies indexing and retrieval into a single sequence-to-sequence framework that directly generates document identifiers (DocIDs) from a user query.
Existing approaches employ diverse identifier designs. 
Pioneering works like DSI~\cite{tay2022transformer} and NCI~\cite{wang2022neural} use structured codes to capture the corpus structure. 
In contrast, methods such as SEAL~\cite{bevilacqua2022autoregressive} and Minder~\cite{li2023multiview} use textual information like titles or keywords, while IncDSI~\cite{kishore2023incdsi} adopts atomic identifiers for direct matching. 
Recent studies have further enhanced retrieval performance by exploring advanced architectures, such as diffusion-based and decoder-only models~\cite{zhao2025diffugr, cai2025exploring}, incorporating learnable DocIDs~\cite{sun2023learning, zeng2024scalable}, and optimizing training strategies, including reinforcement learning and learning-to-rank~\cite{zhou2023enhancing, li2024learning}.

%
Generative retrieval and generative recommendation share a common technical foundation, formulating their respective tasks as conditional generation over discrete identifier spaces. 
However, their underlying logic diverges fundamentally: retrieval aligns queries with documents via semantic matching, whereas recommendation relies on behavioral prediction, synthesizing interaction histories to capture evolving preferences and collaborative signals.

This distinction becomes critical in dynamic environments. 
While recent work~\cite{zhang2025replication} provides a comprehensive evaluation of generative retrieval on dynamic corpora, showing that models can handle new documents without full retraining, these findings do not directly translate to recommendation. 
In retrieval, semantic content often bridges the gap for new documents. 
In contrast, recommendation systems depend heavily on interaction patterns, which are inherently absent in cold-start scenarios. 
Consequently, semantic features alone cannot substitute for the collaborative signals required to model preferences, making the cold-start problem in generative recommendation fundamentally more challenging.

\subsection{Generative recommendation} 
Recent advances have reformulated recommendation as a generative task, where models directly generate item identifiers conditioned on user interaction history and context~\cite{deng2025onerec,zhao2025model,liu2024multi}. 
We organize recent generative recommenders along three key dimensions: identifier design, model architecture, and training objectives.

\begin{enumerate*}[label=(\roman*)]
\item \textit{Identifier design.}  
Early models rely on atomic or random item IDs (RID), which offer no semantic signal and limit generalization to unseen items~\cite{hua2023index}. 
Later work introduces textual identifiers (TID) such as titles or descriptions, which align better with the model’s pre-trained knowledge but may suffer from semantic ambiguity due to token overlap across unrelated items~\cite{tan2024idgenrec}. 
Recent models propose learned semantic codebooks: TIGER and LC-Rec encode each item as a compact sequence of discrete tokens that reflect both semantic and collaborative structure, enabling more precise and generalizable decoding~\cite{rajput2023recommender, zheng2024adapting}.
\item \textit{Model architecture.}  
Most existing methods adopt either encoder-decoder or decoder-only architectures. 
Encoder-decoder models (e.g., P5, TIGER) encode user context and generate item identifiers token by token~\cite{geng2022recommendation}. 
Decoder-only models (e.g., GenRec, BIGRec) treat recommendation as language generation and benefit from alignment with large-scale pre-trained LLMs~\cite{ji2024genrec}. 
DiffGRM introduces an alternative architecture based on diffusion processes, generating identifier codes in parallel via masked denoising, which improves generation consistency and modeling flexibility~\cite{liu2025diffgrm}.
\item \textit{Training objectives.}  
The majority of generative recommenders are trained with supervised fine-tuning (SFT) to imitate observed user-item sequences. 
Beyond SFT, recent models incorporate reinforcement learning or preference alignment to better optimize long-term user satisfaction or ranking utility~\cite{liu2025onerec}. 
For example, OneRec applies direct preference optimization (DPO) to align generation behavior with user-preferred outcomes~\cite{deng2025onerec}.
\end{enumerate*}

In summary, generative recommendation has progressed along multiple dimensions, which is motivated by the need to incorporate pre-trained knowledge and enable semantic generalization. 
These characteristics make generative approaches potentially promising for cold-start scenarios, where interaction signals are limited and inductive generalization is essential.
However, despite the rapid progress of generative recommenders, their effectiveness under cold-start conditions has not been systematically assessed. 
We fill this gap by conducting a comprehensive cold-start evaluation of representative generative recommenders under unified protocols.

\subsection{Cold-start recommendation} 
The cold-start problem, which refers to making recommendations for new items or users with limited interaction history, remains a core challenge for recommender systems~\cite{yuan2023user,du2022metakg,xu2024cmclrec}.
For item cold-start, common approaches include using side information (e.g., titles or attributes)~\cite{monteil2024marec}, transferring knowledge via meta-learning~\cite{vartak2017meta}, and augmenting data through generative models such as conditional VAEs~\cite{zhang2022diverse}. 
For user cold-start, strategies range from using user attributes and clustering to employing bandit-based exploration~\cite{de2025active,nguyen2014cold}.
In the absence of profiles, meta-learning and few-shot adaptation are often used to infer preferences from sparse behavior~\cite{pan2022multimodal}.

More recently, large language models (LLMs) have been introduced into cold-start settings, either to simulate interactions for data augmentation, generate semantically meaningful item identifiers, or directly generate recommendations by prompting the LLM with a cold user’s interaction history or profile to infer preferences and produce relevant items~\cite{wang2024large}.
These generative approaches use pre-trained knowledge and semantic representations to generalize beyond observed interactions, making them potentially suitable for cold-start scenarios.
However, existing evaluations in this space are fragmented, often confounded by inconsistent protocols and entangled design choices. 
These observations motivate the need for a unified and reproducible framework to systematically evaluate generative recommenders under both item and user cold-start conditions, which is exactly what we do in this work.

\section{Problem Formulation}
\label{sec:problem-formulation}

\subsection{Generative recommendation}  
In generative recommendation, we formalize recommendation as a generative task, where the model autoregressively predicts the tokenized identifier of the next item conditioned on a user's interaction history. 
Let $\mathcal{U}$ and $\mathcal{I}$ denote the sets of users and items, respectively. 
Each item $i \in \mathcal{I}$ is associated with a unique identifier $y_i$, represented as a sequence of tokens $[y_i^{(1)}, y_i^{(2)}, \dots, y_i^{(|y_i|)}]$. The format of $y_i$ varies across specific implementations, such as atomic IDs, titles, or semantic codes.

Formally, given a user $u \in \mathcal{U}$ and their interaction history $\mathcal{H}_u = [i_1, i_2, \dots, i_k]$, the model is optimized to maximize the conditional likelihood of the target item's identifier $y_{k+1}$. 
The generative loss $\mathcal{L}_{\text{gen}}$ is defined as the negative log-likelihood of the identifier sequence, calculated using the chain rule of probability:
\begin{equation}
    \mathcal{L}_{\text{gen}} = - \sum_{t=1}^{|y_{k+1}|} \log P(y_{k+1}^{(t)} \mid y_{k+1}^{(<t)}, \mathcal{H}_u; \Theta),
\end{equation}
where $y_{k+1}^{(<t)}$ denotes the prefix of the identifier for item $i_{k+1}$, and $\Theta$ represents the model parameters.

During inference, the recommender system aims to retrieve items that maximize the generation probability. Specifically, for a given user context $\mathcal{H}_u$, the predicted item $\hat{i}$ is identified as:
\begin{equation}
    \hat{i} = \arg\max_{j \in \mathcal{I}} P(y_j \mid \mathcal{H}_u).
\end{equation}
In practice, the top-$K$ recommendation list is generated by ranking candidate items in $\mathcal{I}$ based on their total sequence log-likelihood. To ensure the generation of valid identifiers, constrained decoding (e.g., using a prefix tree or trie) is typically employed to restrict the search space to the set of existing items in $\mathcal{I}$~\cite{rajput2023recommender}.

We study two cold-start scenarios where entities in the test set are unseen during training: item cold-start and user cold-start.

\subsection{Item cold-start}
Let $\mathcal{I}_{tr} \subset \mathcal{I}$ be the set of items observed during training, and $\mathcal{I}_{cold} = \mathcal{I} \setminus \mathcal{I}_{tr}$ denote the set of new items such that $\mathcal{I}_{cold} \cap \mathcal{I}_{tr} = \emptyset$. 
Each cold item $i_c \in \mathcal{I}_{cold}$ is associated with a set of semantic features $\mathbf{x}_{i_c}$, from which its identifier tokens $y_{i_c}$ are constructed. 
At inference time, given a user $u \in \mathcal{U}_{tr}$ and their history $\mathcal{H}_u$, the model is required to score items from the full catalog $\mathcal{I} = \mathcal{I}_{tr} \cup \mathcal{I}_{cold}$. 
The evaluation focuses on the ranking performance specifically for items in $\mathcal{I}_{cold}$, testing the model's ability to generalize to identifiers not seen during the optimization of $\mathcal{L}_{\text{gen}}$.

\subsection{User cold-start}
Let $\mathcal{U}_{tr} \subset \mathcal{U}$ be the set of users observed during training, and $\mathcal{U}_{cold} = \mathcal{U} \setminus \mathcal{U}_{tr}$ denote the set of new users such that $\mathcal{U}_{cold} \cap \mathcal{U}_{tr} = \emptyset$. 
For each cold user $u_c \in \mathcal{U}_{cold}$, the model is provided with a limited interaction history $\mathcal{H}_{u_c} = [i_1, i_2, \dots, i_m]$ consisting of items from $\mathcal{I}_{tr}$. 
Unlike users in $\mathcal{U}_{tr}$ whose interaction patterns were seen during training, users in $\mathcal{U}_{cold}$ are potentially out-of-distribution in terms of their global identity. 
At inference time, the model must predict the next item $j \in \mathcal{I}$ by conditioning solely on the provided sequence $\mathcal{H}_{u_c}$. 
The evaluation measures the model's performance in generalizing to the behavioral contexts of new users without any prior exposure to their specific interaction histories.
\section{Experimental Setup}
\label{sec:setup}

This section describes the datasets, reproduced methods, and evaluation protocol used to systematically analyze the cold-start behavior of generative recommenders. 

\subsection{Datasets}
\label{sec:data}

We evaluate our proposed framework on three large-scale benchmarks commonly used in the recommendation literature: Amazon-Toys, MicroLens, and Steam. 
These datasets are selected for their high-quality item metadata and diverse interaction patterns, providing a robust testbed for generative models in cold-start scenarios.
\begin{enumerate*}[label=(\roman*)]
\item \textbf{Amazon-Toys.}\footnote{\url{https://amazon-reviews-2023.github.io}.}
This dataset is a subset of the \textit{Toys and Games} category from the Amazon Review corpus. It contains dense e-commerce interactions and comprehensive item metadata, including product titles and hierarchical category paths.
\item \textbf{MicroLens.}\footnote{\url{https://github.com/westlake-repl/MicroLens}.}  
A large-scale micro-video recommendation dataset collected from a real-world platform, designed for content-driven recommendation at scale.  
It provides rich multi-modal item-side information (e.g., video titles, cover images, audio, and full videos).
\item \textbf{Steam.}\footnote{\url{https://github.com/kang205/SASRec}.}  
A game recommendation dataset collected from the Steam platform, reflecting diverse user behaviors with varying playtime and preferences.  
Games are annotated with textual descriptions, developer tags, and user-defined categories, offering rich side information to support cold-start generalization.
\end{enumerate*}
We select these datasets because they provide large-scale interactions with high-quality item-side metadata that is essential for studying generative recommenders under cold-start settings, whereas many commonly used datasets lack reliable metadata or are too small to support controlled cold-start evaluation.
The dataset statistics are shown in Table~\ref{tab:dataset_stats}.

\begin{table}[htbp]
\centering
\caption{Statistics of the datasets used in the experiments. ``W-'' and ``C-'' denote Warm and Cold, respectively. ``Inter.'' stands for the number of interactions.}
\label{tab:dataset_stats}
\begin{adjustbox}{width=\linewidth}
\begin{tabular}{lccccc}
\toprule
\textbf{Dataset} & \textbf{W-User} & \textbf{C-User} & \textbf{W-Item} & \textbf{C-Item} & \textbf{Inter.} \\
\midrule
Amazon-Toys & 17,479 & 1,942 & 11,876 & \phantom{0}48 & 148,185 \\
MicroLens   & 45,000 & 5,000 & 18,490 & 730 & 359,709 \\
Steam       & 36,000 & 4,000 & 10,963 & 635 & 276,281 \\
\bottomrule
\end{tabular}
\end{adjustbox}
\end{table}

To ensure data quality and simulate realistic cold-start scenarios, we perform several pre-processing steps.
Firstly, we apply a 5-core filtering to all datasets, removing users and items with fewer than five interactions. 
To evaluate the \textbf{item cold-start} performance, we adopt the standard leave-one-out protocol, where the last item in each user sequence is used for testing, the second-to-last item for validation, and the remaining interactions for training. 
Interactions in the validation and test sets are further categorized according to whether their target items appear in the training set: interactions involving items seen during training are treated as warm-start cases, while those involving previously unseen items are treated as cold-start cases. 

For the user cold-start scenario, we randomly hold out 10\% of the users and exclude all their interactions from the training set. 
We then evaluate these users under the same leave-one-out protocol. 
To simulate the sparse interaction history of new users, we randomly retain 1 to 10 historical interactions for each cold-start user as the observed context, and use the subsequent interactions for validation and testing. 
For the remaining users, all available training interactions are retained, and they are treated as warm-start users.

\subsection{Methods reproduced}
\label{sec:baseline}
\begin{table}[t]
\small
\centering
\setlength{\tabcolsep}{4pt}
\caption{Comparison of generative recommendation methods. Enc--Dec denotes encoder--decoder architecture, and Dec denotes decoder-only architecture.}
\label{tab:methods}
\begin{tabular}{l r l l l}
\toprule
\textbf{Model} & \multicolumn{1}{c}{\textbf{Arch.}} & \textbf{ID Type} & \textbf{Codebook} & \textbf{Training} \\
\midrule
RID~\cite{hua2023index}      & Enc--Dec & Atomic   & --              & SFT \\
TID~\cite{hua2023index}      & Enc--Dec & Textual  & --              & SFT \\
P5~\cite{geng2022recommendation} & Enc--Dec & Atomic   & --              & SFT \\
GenRec~\cite{ji2024genrec}   & Dec      & Textual  & --              & SFT \\
TIGER~\cite{rajput2023recommender} & Enc--Dec & Semantic & RQ-VAE         & SFT \\
LC-Rec~\cite{zheng2024adapting}    & Dec      & Semantic & RQ-VAE         & SFT \\
OneRec~\cite{deng2025onerec} & Dec & Semantic & Balanced-KMeans & SFT+RL \\
DiffGRM~\cite{liu2025diffgrm} & Enc--Dec & Semantic & OPQ            & SFT \\
\bottomrule
\end{tabular}
\end{table}

To provide a comprehensive evaluation, we reproduce a diverse set of representative baselines in this study. These methods span both traditional sequential models and modern generative models, allowing for a rigorous comparison across different recommendation paradigms. 

\subsubsection{Traditional baselines}
We include two classical sequential recommendation models to provide a performance reference for evaluating the generative approaches.
\begin{enumerate*}[label=(\roman*)]
\item \textbf{SASRec}~\cite{kang2018self}: A transformer-based recommender model that models user behavior as a sequence and uses self-attention to capture long-range dependencies.  
\item \textbf{GRU4Rec}~\cite{hidasi2015session}: A GRU-based recommender model that processes user interaction sequences and predicts the next item via recurrent hidden states.
\end{enumerate*}

\subsubsection{Generative baselines}
We reproduce eight generative models that represent the current state-of-the-art in item identifier design and training objectives:
\begin{enumerate*}[label=(\roman*)]
    \item \textbf{RID}~\cite{hua2023index}: Uses random numeric item IDs tokenized into subword units, offering simple LLM compatibility but no semantic structure.
    \item \textbf{TID}~\cite{hua2023index}: Represents items with their textual titles, using semantic content but introducing potential noise from overlapping tokens.
    \item \textbf{P5}~\cite{geng2022recommendation}: A T5-based encoder-decoder that reformulates recommendation as text-to-text generation using prompt templates and atomic item IDs.
    \item \textbf{GenRec}~\cite{ji2024genrec}: A decoder-only LLM trained to generate item names from user history using textual prompts, without candidate scoring.
    \item \textbf{TIGER}~\cite{rajput2023recommender}: Encodes items as discrete semantic codes and trains an encoder-decoder to generate these codes autoregressively from user history.
    \item \textbf{LC-Rec}~\cite{zheng2024adapting}: Learns vector-quantized item codes and fine-tunes a decoder-only LLM to align language and collaborative semantics.
    \item \textbf{OneRec}~\cite{deng2025onerec}: A production-scale encoder-decoder trained with both SFT and reinforcement learning to generate learned semantic item codes.
    \footnote{As the original implementation of OneRec is not publicly available, we instantiate the framework using MiniOneRec~\cite{kong2025minionerec}, a publicly available and reproducible implementation that fully reproduces the model and training methodology described in the original OneRec paper, including semantic-ID generation, supervised fine-tuning, and preference alignment.}
    \item \textbf{DiffGRM}~\cite{liu2025diffgrm}: A diffusion-based model that generates item codes in parallel using masked denoising, avoiding autoregressive constraints.
\end{enumerate*}

Table~\ref{tab:methods} summarizes the key characteristics of each reproduced generative model, including identifier type and training strategy, which are central factors in our analysis.

\subsection{Evaluation and implementation details}
To comprehensively evaluate cold-start behavior, we conduct experiments under three distinct evaluation protocols. 
These settings are designed to isolate the model's performance on observed data versus its ability to handle unseen entities:
\begin{itemize}
    \item \textbf{Warm-start setting}: This serves as the baseline setting, where the goal is to recommend previously observed items to existing users.
    \item \textbf{User cold-start setting}: We evaluate the model's performance in recommending items to new users based solely on their short-term behavioral context.
    \item \textbf{Item cold-start setting}: We assess the model's ability to surface new items that were not present in the training set.
\end{itemize}

\subsubsection{Evaluation metrics.}
For each setting, we report the top-$K$ recommendation performance using two standard ranking metrics: \textbf{Recall@$K$} and \textbf{NDCG@$K$}. 
Following common practice in the literature, we set $K=10$ as the primary threshold for comparison. 
All metrics are averaged across all test users to ensure a representative measure of global performance.

\subsubsection{Implementation details.}

We implement the traditional sequential baselines, SASRec and GRU4Rec, using the RecBole library,\footnote{\url{https://github.com/RUCAIBox/RecBole}} which provides a standardized and optimized environment for these models. 
Experiments are conducted on 8 NVIDIA A800 GPUs, with distributed training enabled for GR methods to accommodate their large parameter sizes.
For the generative recommendation methods, we use their respective official implementations and strictly adhere to the default hyper-parameter configurations reported in the original literature to ensure rigorous reproducibility. 
The construction of training sequences follows the specific data augmentation and formatting strategies of each method, such as sliding window partitioning~\cite{liu2025onerec} and random context length sampling~\cite{liu2025diffgrm}, to maintain consistency with their original design. 
All models are evaluated under the uniform protocol described above to ensure a fair and equitable comparison across different recommendation paradigms.

\section{Experiments}
\label{sec:exp}

\begin{table*}[t]
\centering
\caption{User cold-start recommendation performance comparison. Performance is measured using Recall@10 (R@10) and NDCG@10 (N@10), with the best and second-best results in each category bolded and underlined, respectively. 
The best performing models achieve significant improvements over the strongest competing baseline (paired t-test, p<0.05).}
\label{tab:user_cold_start_result}
\begin{tabular}{l cccc cccc cccc} 
\toprule
\multirow{3}{*}{\textbf{Model}} & \multicolumn{4}{c}{\textbf{Amazon-Toys}} & \multicolumn{4}{c}{\textbf{MicroLens}} & \multicolumn{4}{c}{\textbf{Steam}} \\
\cmidrule(lr){2-5} \cmidrule(lr){6-9} \cmidrule(lr){10-13}
& \multicolumn{2}{c}{Warm-test} & \multicolumn{2}{c}{Cold-test} & \multicolumn{2}{c}{Warm-test} & \multicolumn{2}{c}{Cold-test} & \multicolumn{2}{c}{Warm-test} & \multicolumn{2}{c}{Cold-test} \\
\cmidrule(lr){2-3} \cmidrule(lr){4-5} \cmidrule(lr){6-7} \cmidrule(lr){8-9} \cmidrule(lr){10-11} \cmidrule(lr){12-13}
& R@10 & N@10 & R@10 & N@10 & R@10 & N@10 & R@10 & N@10 & R@10 & N@10 & R@10 & N@10 \\
\midrule
\multicolumn{13}{l}{\cellcolor{gray!10}\textit{\textbf{Traditional baselines}}} \\
SASRec  & 0.0672 & 0.0381 & 0.0418 & 0.0211 & 0.0836 & 0.0475 & 0.0506 & 0.0264 & 0.0764 & 0.0411 & 0.0533 & 0.0301 \\
GRU4Rec & 0.0171 & 0.0090 & 0.0112 & 0.0058 & 0.0708 & 0.0406 & 0.0103 & 0.0034 & 0.0667 & 0.0376 & 0.0413 & 0.0263 \\
\midrule
\multicolumn{13}{l}{\cellcolor{gray!10}\textit{\textbf{Generative baselines}}} \\
RID     & 0.0518 & 0.0346 & 0.0239 & 0.0126 & 0.0508 & 0.0261 & 0.0232 & 0.0117 & 0.0412 & 0.0243 & 0.0256 & 0.0178 \\
TID     & 0.0414 & 0.0251 & 0.0335 & 0.0175 & 0.0464 & 0.0274 & 0.0372 & 0.0221 & 0.0475 & 0.0285 & 0.0362 & 0.0229 \\
P5      & 0.0430 & 0.0257 & 0.0322 & 0.0179 & 0.0488 & 0.0303 & 0.0345 & 0.0209 & 0.0303 & 0.0171 & 0.0250 & 0.0174 \\
GenRec  & 0.0253 & 0.0158 & 0.0196 & 0.0115 & 0.0346 & 0.0186 & 0.0256 & 0.0168 & 0.0292 & 0.0181 & 0.0247 & 0.0171 \\
TIGER   & 0.0654 & 0.0353 & 0.0468 & 0.0259 & 0.0710 & 0.0372 & 0.0554 & 0.0296 & 0.0721 & 0.0395 & 0.0527 & 0.0293 \\
LC-Rec  & 0.0753 & 0.0395 & 0.0576 & 0.0299 & 0.0822 & 0.0427 & 0.0603 & 0.0346 & \underline{0.0812} & \underline{0.0518} & \underline{0.0604} & \underline{0.0371} \\
OneRec  & \underline{0.0804} & \underline{0.0451} & \textbf{0.0602} & \textbf{0.0316} & \textbf{0.0841} & \textbf{0.0485} & \textbf{0.0629} & \textbf{0.0372} & \textbf{0.0831} & \textbf{0.0533} & \textbf{0.0609} & \textbf{0.0377} \\
DiffGRM & \textbf{0.0818} & \textbf{0.0463} & \underline{0.0595} & \underline{0.0307} & 0.0833 & \underline{0.0477} & \underline{0.0614} & \underline{0.0362} & 0.0796 & 0.0498 & 0.0583 & 0.0352 \\
\bottomrule
\end{tabular}
\end{table*}
\begin{table*}[t]
\centering
\caption{Item cold-start recommendation performance comparison. Performance is measured using Recall@10 (R@10) and NDCG@10 (N@10), with the best and second-best results in each category bolded and underlined, respectively. The best performing models achieve significant improvements over the strongest competing baseline (paired t-test, p<0.05).}
\label{tab:item_cold_start_result}
\begin{tabular}{l cccc cccc cccc} 
\toprule
\multirow{3}{*}{\textbf{Model}} & \multicolumn{4}{c}{\textbf{Amazon-Toys}} & \multicolumn{4}{c}{\textbf{MicroLens}} & \multicolumn{4}{c}{\textbf{Steam}} \\
\cmidrule(lr){2-5} \cmidrule(lr){6-9} \cmidrule(lr){10-13}
& \multicolumn{2}{c}{Warm-test} & \multicolumn{2}{c}{Cold-test} & \multicolumn{2}{c}{Warm-test} & \multicolumn{2}{c}{Cold-test} & \multicolumn{2}{c}{Warm-test} & \multicolumn{2}{c}{Cold-test} \\
\cmidrule(lr){2-3} \cmidrule(lr){4-5} \cmidrule(lr){6-7} \cmidrule(lr){8-9} \cmidrule(lr){10-11} \cmidrule(lr){12-13}
& R@10 & N@10 & R@10 & N@10 & R@10 & N@10 & R@10 & N@10 & R@10 & N@10 & R@10 & N@10 \\
\midrule
\multicolumn{13}{l}{\cellcolor{gray!10}\textit{\textbf{Traditional baselines}}} \\
SASRec  & 0.0675 & 0.0374 & 0.0012 & 0.0003 & \underline{0.0841} & \underline{0.0482} & 0.0007 & 0.0001 & 0.0769 & 0.0412 & 0.0015 & 0.0004 \\
GRU4Rec & 0.0174 & 0.0091 & 0.0000 & 0.0000 & 0.0714 & 0.0414 & 0.0000 & 0.0000 & 0.0675 & 0.0386 & 0.0007 & 0.0002 \\
\midrule
\multicolumn{13}{l}{\cellcolor{gray!10}\textit{\textbf{Generative baselines}}} \\
RID     & 0.0522 & 0.0349 & 0.0000 & 0.0000 & 0.0511 & 0.0263 & 0.0000 & 0.0000 & 0.0413 & 0.0243 & 0.0000 & 0.0000 \\
TID     & 0.0416 & 0.0250 & \textbf{0.0312} & \textbf{0.0181} & 0.0467 & 0.0276 & \textbf{0.0121} & \textbf{0.0078} & 0.0476 & 0.0285 & \textbf{0.0214} & \textbf{0.0141} \\
P5      & 0.0431 & 0.0257 & 0.0000 & 0.0000 & 0.0490 & 0.0304 & 0.0000 & 0.0000 & 0.0315 & 0.0179 & 0.0000 & 0.0000 \\
GenRec  & 0.0251 & 0.0157 & \underline{0.0179} & \underline{0.0102} & 0.0349 & 0.0191 & \underline{0.0102} & \underline{0.0067} & 0.0295 & 0.0189 & \underline{0.0196} & \underline{0.0115} \\
TIGER   & 0.0652 & 0.0352 & 0.0034 & 0.0009 & 0.0716 & 0.0377 & 0.0043 & 0.0012 & 0.0723 & 0.0386 & 0.0038 & 0.0015 \\
LC-Rec  & 0.0756 & 0.0396 & 0.0055 & 0.0013 & 0.0823 & 0.0432 & 0.0076 & 0.0031 & \underline{0.0815} & \underline{0.0520} & 0.0079 & 0.0032 \\
OneRec  & \underline{0.0802} & \underline{0.0451} & 0.0043 & 0.0013 & \textbf{0.0844} & \textbf{0.0487} & 0.0078 & 0.0032 & \textbf{0.0834} & \textbf{0.0536} & 0.0069 & 0.0034 \\
DiffGRM & \textbf{0.0827} & \textbf{0.0463} & 0.0022 & 0.0004 & 0.0836 & 0.0479 & 0.0032 & 0.0008 & 0.0801 & 0.0502 & 0.0027 & 0.0005 \\
\bottomrule
\end{tabular}
\end{table*}

Our experiments are designed to answer the research questions introduced in Section~\ref{sec:introduction}. 
We first establish the overall behavior of representative methods under warm-start, user cold-start, and item cold-start settings. 
We then perform a controlled factor-wise analysis to isolate the effects of model scale, identifier design, and training strategy. 
Finally, since identifier design emerges as the most influential factor for item cold-start performance, we conduct a focused follow-up analysis to better understand why semantic identifiers fail on unseen items.

\subsection{Overall performance (RQ1)}
We reproduce all methods on three benchmark datasets described in Section~\ref{sec:data} and evaluate their recommendation performance under three settings, i.e., warm-start, user cold-start, and item cold-start, in Table~\ref{tab:user_cold_start_result} and Table~\ref{tab:item_cold_start_result}.
From these experiments, we derive several key observations.

First, the item cold-start is substantially more challenging than the user cold-start.
Across all datasets and model families, performance under item cold-start conditions drops sharply compared to warm-start, whereas user cold-start exhibits only moderate degradation.
Even strong generative recommenders remain relatively stable under user cold-start, but deteriorate markedly under item cold-start. 
This suggests that generalizing to unseen items is substantially harder than recommending for sparsely observed users.

Second, textual identifiers improve item cold-start performance but introduce a clear trade-off.
Methods using textual identifiers consistently outperform those using atomic or learnable semantic codes in an item cold-start setting, indicating that using pre-trained semantic knowledge is beneficial when interaction signals are unavailable.
However, these same textual models tend to underperform in warm-test and user cold-test settings.
This trade-off suggests that while textual tokens provide stronger semantic generalization for unseen items, they may reduce identifier discriminability or introduce ambiguity among frequently observed items, which harms performance when sufficient interaction data is available.

Third, generation architecture affects robustness in cold-start settings.
Among semantic-code based methods, models that generate identifiers in a more compositional or parallel manner, such as diffusion-based approaches, achieve stronger and more stable performance across settings.
In particular, DiffGRM consistently ranks among the top-performing models across settings, and remains competitive in an item cold-start setting, suggesting that mitigating strong sequential dependencies may enhance robustness when generalizing beyond the training distribution
In contrast, traditional collaborative baselines degrade most severely for item cold-start, as expected, since they primarily rely on interaction signals and lack explicit semantic priors to support generalization to unseen items.


\subsection{Controlled factor analysis}

Having established the overall cold-start asymmetry, we next isolate three key design dimensions that are often entangled in prior work: model scale, identifier design, and training strategy.
This controlled analysis allows us to quantify which factors are most responsible for the observed performance differences under cold-start settings.

\subsubsection{Effect of model scale (RQ2)}

\begin{figure}[htbp]
    \centering
\setlength{\belowcaptionskip}{-0.5cm}
    \includegraphics[width=\linewidth]{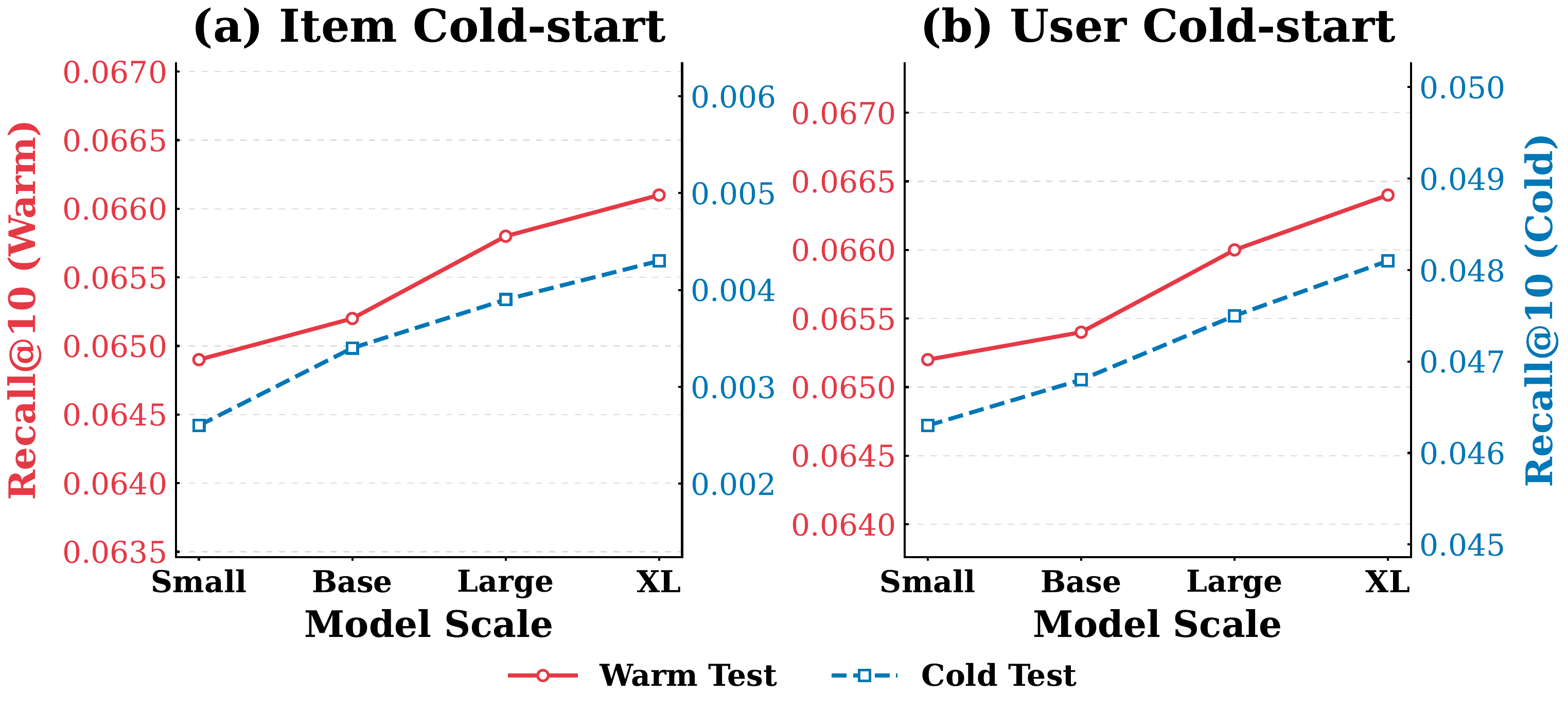} 
    \caption{Effect of model scale on warm-start and cold-start performance. Recall@10 is reported for TIGER with different Flan-T5 backbones under item and user cold-start settings.}
    \label{fig:scaling}
\end{figure}

To isolate the effect of model scale, we replace the TIGER backbone with four Flan-T5 variants\footnote{\url{https://huggingface.co/docs/transformers/en/model_doc/flan-t5},} from \textit{small} to \textit{xl}, while keeping identifier design and training strategy fixed. 
Figure~\ref{fig:scaling} reports the results under warm-start, user cold-start, and item cold-start.
Based on these results, we derive several observations on how model scale affects cold-start behavior in generative recommendation.

First, a scaling trend is clearly observable across all settings.
As model size increases from \textit{flan-t5-small} to \textit{flan-t5-xl}, performance consistently improves under warm-start, user cold-start, and item cold-start.
This confirms that larger PLMs provide stronger representation capacity and better semantic modeling ability, which positively influences recommendation quality under cold-start setting.

Second, the magnitude of scaling gains under cold-start settings remains limited.
Although performance increases monotonically with model size, the relative improvements under item cold-start settings are modest and the absolute performance gap between warm and cold settings remains substantial.
Even the largest model does not close the cold-start gap, indicating that scaling alone cannot fundamentally address the generalization challenge posed by cold users and items.


\subsubsection{Effect of identifier design (RQ3)}
Identifier design plays a central role in generative recommendation, as items are predicted through discrete tokens that serve as the model’s output space. 
To isolate the effect of identifier design, we use TIGER and vary only the item identifier while fixing the architecture, backbone, and training strategy. 
We compare atomic IDs, textual titles, and semantic codes constructed by RQ-VAE, Balanced k-means, and OPQ.



\begin{figure}[t]
    \centering
    \includegraphics[width=\linewidth]{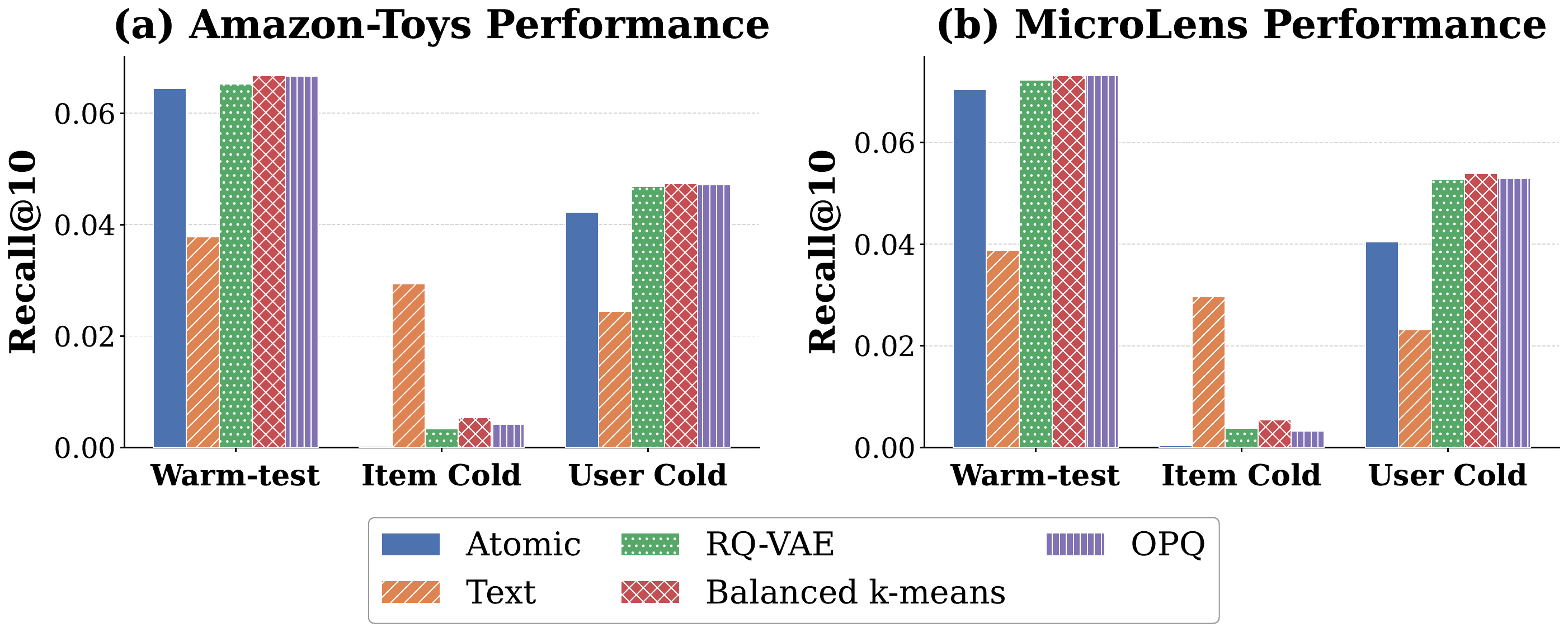} 
\caption{Effect of identifier design on warm-start and cold-start performance. Recall@10 is reported for atomic IDs, textual titles, and semantic IDs under warm-start, item cold-start, and user cold-start settings.}
    \label{fig:id_design}
\end{figure}
We report the recommendation performance of these identifier variants under warm-start, user cold-start, and item cold-start settings in Figure~\ref{fig:id_design}.
From these results, we derive several observations regarding how identifier design affects cold-start generalization in generative recommendation.


First, identifier design induces a clear trade-off between warm-start accuracy and item cold-start generalization.
Textual identifiers achieve the strongest performance under item cold-start, showing the benefit of exposing semantic cues from pre-training when the target items are unseen.
However, the same textual identifiers substantially underperform in warm-test and user cold-start.
This suggests that while titles provide transferable semantics, they are not optimized to be uniquely discriminative identifiers, and token overlap or semantic ambiguity can increase decoding confusion when sufficient behavioral signals exist.

Second, semantic-code identifiers are robust for user cold-start but less stable under item cold-start.
Across settings, quantized semantic codes maintain strong performance in warm-test and achieve the best results under user cold-start, indicating that a compact and structured output space can be reliably generated and effectively exploited when the candidate items are largely within the training distribution.
In contrast, item cold-start performance drops sharply for several semantic-code designs.

Third, purely atomic identifiers fail most severely under item cold-start settings due to an output-space mismatch.
With atomic IDs, cold items correspond to unseen identifier tokens during training, so the model has little opportunity to learn to generate them.
As a result, item cold-start performance collapses even when warm-test remains competitive.
This highlights a fundamental limitation of atomic-ID generation for cold item recommendation.



\subsubsection{Effect of training strategies (RQ4)}



Since some recent generative recommendation methods further adopt reinforcement learning (RL) on top of supervised fine-tuning (SFT), we examine whether this additional optimization improves cold-start performance. 
We compare representative RL-based methods before and after the RL stage, while keeping the model architecture and identifier design fixed. 
Table~\ref{tab:rl_comparison_final} reports the results under warm-start, user cold-start, and item cold-start.

\begin{table}[t]
\centering
\caption{Performance impact of RL on Amazon-Toys measured by Recall@10. The \textcolor{red}{red numbers} in parentheses represent the percentage change ($\Delta$) relative to the SFT baseline.}
\label{tab:rl_comparison_final}
\setlength{\tabcolsep}{6pt} 
\begin{tabular}{l l l l}
\toprule
\textbf{Scenario} & \textbf{Method} & \multicolumn{1}{c}{\textbf{Warm-test}} & \textbf{Cold-test} \\
\midrule
\multirow{5}{*}{\makecell[l]{\textbf{Item}\\\textbf{Cold-start}}} 
 & OneRec (SFT) & 0.0804 & 0.0045 \\
 & \quad + RL & 0.0802 {\scriptsize(\textcolor{red}{-0.2})} & 0.0043 {\scriptsize(\textcolor{red}{-4.4})} \\
\cmidrule(lr){2-4}
 & TIGER (SFT)  & 0.0652 & 0.0034 \\
 & \quad + RL & 0.0643 {\scriptsize(\textcolor{red}{-1.3})} & 0.0031 {\scriptsize(\textcolor{red}{-8.8})} \\
\midrule
\multirow{5}{*}{\makecell[l]{\textbf{User}\\\textbf{Cold-start}}} 
 & OneRec (SFT) & 0.0806 & 0.0605 \\
 & \quad + RL & 0.0804 {\scriptsize(\textcolor{red}{-0.2})} & 0.0602 {\scriptsize(\textcolor{red}{-0.5})} \\
\cmidrule(lr){2-4}
 & TIGER (SFT)  & 0.0654 & 0.0468 \\
 & \quad + RL & 0.0641 {\scriptsize(\textcolor{red}{-1.9})} & 0.0459 {\scriptsize(\textcolor{red}{-1.9})} \\
\bottomrule
\end{tabular}
\end{table}

From these results, we observe that adding an RL stage on top of SFT does not consistently improve performance under cold-start settings. 
In several cases, RL yields only marginal changes and can even slightly degrade performance under both user and item cold-start. 
This suggests that improved alignment with training-time reward signals does not automatically translate into better robustness under distribution shift.

A possible explanation is that RL optimization is driven by reward signals constructed from the training distribution, which may encourage the model to exploit frequent identifier patterns and in-distribution item structures. 
Under cold-start conditions, especially for unseen items, the model must extrapolate beyond these patterns. 
In this regime, optimizing for reward on seen data may not transfer effectively, thereby limiting generalization.

Overall, these findings indicate that reinforcement learning is non-trivial to tune for cold-start scenarios and that cold-start robustness likely requires reward design and training objectives that explicitly account for distribution shift rather than relying solely on post-hoc alignment.


\subsection{Semantic ID failure in item cold-start}

\begin{figure}[t]
    \centering
    \includegraphics[width=\linewidth]{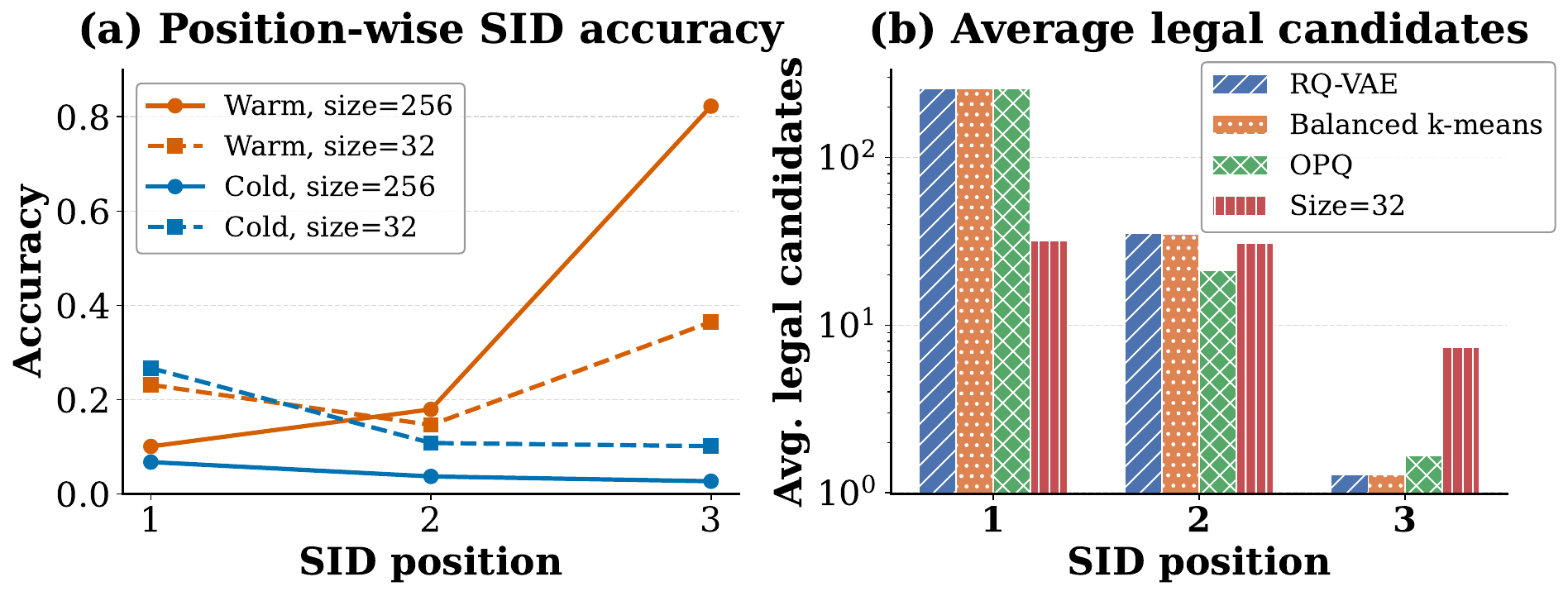}
\caption{Decoding behavior of semantic identifiers under item cold-start. (a) Position-wise SID prediction accuracy with different codebook sizes. (b) Average legal candidates at each SID position under constrained decoding.}
    \label{fig:sid_analysis}
\end{figure}

As a follow-up to RQ3, we further analyze why semantic identifiers remain weak under item cold-start despite their overall robustness.
This is surprising because semantic IDs encode item-side information that should, in principle, support unseen-item generalization. We therefore analyze their decoding behavior in Figure~\ref{fig:sid_analysis}. In generative recommendation, recommending a cold item requires generating its full identifier sequence. Since semantic IDs are typically generated with trie-constrained decoding, each prefix restricts the legal next tokens. Thus, item cold-start depends not only on whether a cold item has a meaningful identifier, but also on whether the model can reach the correct identifier branch during decoding.

Figure~\ref{fig:sid_analysis}(a) shows opposite position-wise trends for warm and cold items. Under warm-start, SID prediction accuracy increases from 0.1008 to 0.1794 and 0.8225, indicating that later tokens become easier once the model enters the correct prefix branch. 
Under item cold-start, however, accuracy decreases from 0.0676 to 0.0372 and 0.0270, suggesting that early decoding errors are amplified rather than corrected. Figure~\ref{fig:sid_analysis}(b) explains this pattern: under trie-constrained decoding, the legal candidate space shrinks sharply. 
For RQ-VAE-based semantic IDs, the number of legal next-token candidates drops from 256.0 at the first position to 35.25 at the second and only 1.28 at the third, with similar contractions observed for Balanced k-means and OPQ.
This contraction helps warm items by narrowing the search along learned training branches, but hurts cold items because once decoding enters a frequent seen branch, the model has little room to recover the unseen identifier.

\begin{table}[htbp]
\centering
\caption{Effect of SID codebook size on warm-start and item cold-start performance. Smaller codebooks improve cold-start retrieval but reduce warm-start accuracy.}
\label{tab:codebook_ablation}
\begin{tabular}{lccc}
\toprule
Setting & Metric & Baseline & Codebook size = 32 \\
\midrule
\multirow{2}{*}{Warm-start}
& Recall@10 & 0.0652 & 0.0601 \\
& NDCG@10   & 0.0352 & 0.0329 \\
\midrule
\multirow{2}{*}{Cold-start}
& Recall@10 & 0.0034 & 0.0068 \\
& NDCG@10   & 0.0009 & 0.0023 \\
\bottomrule
\end{tabular}
\end{table}

We further verify this explanation by reducing SID codebook size from 256 to 32, which makes the legal candidate space more balanced across positions.
As shown in Table~\ref{tab:codebook_ablation}, this improves item cold-start Recall@10 from 0.0034 to 0.0068. However, warm-start performance drops from 0.0652 to 0.0601 in Recall@10. 
This confirms a decoding-space trade-off: smaller codebooks reduce branch sparsity and improve cold-item recoverability, but weaken the discriminative power needed for warm-start recommendation. 

Overall, semantic IDs fail on item cold-start partly because their constrained decoding space amplifies training-observed branch bias. 
Therefore, cold-start-oriented identifier design should consider not only semantic quality, but also branch balance and decoding recoverability. 

Building on this, a critical direction for future work is to optimize encoding structures by ensuring a more uniform distribution of legal candidates across decoding positions, which directly alleviates the candidate collapse problem. 
Meanwhile, future designs also should prevent this structural balancing from degrading warm-start accuracy, ultimately ensuring robustness for cold items while fully preserving discriminative power for seen ones.

\section{Conclusion}
\label{sec:conclusion}

In this work, we revisited the common assumption that generative recommendation built on pre-trained language models can naturally alleviate cold-start challenges. This assumption remains difficult to verify because existing studies often lack unified cold-start protocols and entangle key design factors, including model scale, identifier design, and training strategy.
We addressed this gap through a systematic reproducibility study under unified user and item cold-start settings. By reproducing representative generative recommenders and isolating key design dimensions, we provide a controlled analysis of their cold-start behavior. 

Our results show that:
\begin{enumerate*}[label=(\roman*)]
\item cold-start is highly asymmetric, with item cold-start being much harder than user cold-start;
\item scaling the backbone PLM brings consistent but limited gains and does not close the cold-start gap;
\item identifier design is the dominant factor: textual identifiers improve item cold-start but hurt warm-start and user cold-start performance, while semantic identifiers are more robust overall but suffer from constrained decoding bias on unseen items;
and
\item reinforcement learning does not consistently improve cold-start performance and may reduce robustness under distribution shift.
\end{enumerate*}

Overall, our findings clarify when generative recommenders help cold-start scenarios and which design choices matter most for generalization to unseen users and items. 
To overcome the specific decoding bottlenecks of semantic identifiers, future research should explore improved encoding structures that alleviate the legal candidate collapse problem. 
Resolving this structural limitation will actively mitigate the generation bias toward seen IDs without compromising overall warm-start performance. 
We hope our unified protocols and factor-wise evidence support more rigorous future research on cold-start generative recommendation.

\subsubsection*{\rm\textbf{Limitations}}
Despite providing a systematic and controlled reproducibility study, our analysis does not fully capture the complexity of real-world recommendation scenarios, which often involve richer modalities, dynamic item pools, and evolving user preferences.
Moreover, although we analyze how identifier design affects cold-start robustness through decoding-level empirical studies, we do not provide a formal theoretical account of the underlying mechanisms, such as how quantization structure or token factorization shape distribution shift and decoding stability.


\section*{Resources}
We share the following resources to facilitate reproducibility of our work (see Footnote~\ref{fn:1}): 
\begin{enumerate*}[label=(\roman*)]
    \item source code for the reproduced models and evaluation pipeline;
    \item scripts for dataset preprocessing and cold-start split construction; and
    \item configuration files and documentation required to reproduce all experiments reported in this paper.
\end{enumerate*}

\begin{acks}
    This research was (partially) supported by the Dutch Research Council (NWO), under project numbers 024.004.022, NWA.1389.20.\-183, and KICH3.LTP.20.006, and the European Union under grant agreement No.\ 101201510 (UNITE).
    This research was also supported by the Shandong Provincial Key Laboratory of Computing-Network Integration and the Qingdao Key Laboratory of Trustworthy Artificial Intelligence, Shandong University.

    Views and opinions expressed are those of the author(s) only and do not necessarily reflect those of their respective employers, funders and/or granting authorities.
\end{acks}

\bibliographystyle{ACM-Reference-Format}
\balance
\bibliography{bibfile}

\end{document}